\documentclass[referee]{aa}
\usepackage{graphicx}
\usepackage[varg]{txfonts}
\usepackage{cases}

\begin{document}
   \title{A swirling flare-related EUV jet}

   \author{Q. M. Zhang \and H. S. Ji}

   \institute{Key Laboratory for Dark Matter and Space Science, Purple
              Mountain Observatory, CAS, Nanjing 210008, China \\
              \email{zhangqm@pmo.ac.cn}
              }

   \date{Received; accepted}
    \titlerunning{A flare-related EUV jet}
    \authorrunning{Q. M. Zhang \& H. S. Ji}

  \abstract
   {}
   {We report our observations of a swirling flare-related EUV jet on 
   2011 October 15 at the edge of NOAA active region 11314.}
   {We utilised the multiwavelength 
    observations in the extreme-ultraviolet (EUV) passbands
    from the Atmospheric Imaging Assembly (AIA) aboard the Solar 
    Dynamics Observatory (SDO). We extracted a wide slit along the jet
    axis and 12 thin slits across its axis to investigate the longitudinal motion
    and transverse rotation. We also used data from the Extreme-Ultraviolet 
    Imager (EUVI) aboard the Solar TErrestrial RElations Observatory 
    (STEREO) spacecraft to investigate the three-dimensional (3D) structure 
    of the jet. Gound-based H$\alpha$ images from the El Teide as a 
    member of the Global Oscillation Network Group (GONG) provide a good
    opportunity to explore the relationship between the cool surge and hot jet.
    Line-of-sight magnetograms from the Helioseismic and Magnetic Imager 
    (HMI) aboard SDO enable us to study the magnetic evolution of the flare/jet
    event. We carried out potential-field extrapolation to figure out the magnetic
    configuration associated with the jet.}
   {The onset of jet eruption coincided with the start time of C1.6 flare impulsive phase. 
   The initial velocity and acceleration of the longitudinal motion were 254$\pm10$ km s$^{-1}$
    and $-97\pm5$ m s$^{-2}$, respectively. The jet presented helical structure 
    and transverse swirling motion at the beginning of its eruption. The counter-clockwise 
    rotation slowed down from an average velocity of $\sim$122 km s$^{-1}$ to 
    $\sim$80 km s$^{-1}$. The interwinding thick threads of the jet untwisted
    into multiple thin threads during the rotation that lasted for 1 cycle with 
    a period of $\sim$7 min and an amplitude that increases from 
    $\sim$3.2 Mm at the bottom to $\sim$11 Mm at the upper part.
    Afterwards, the curtain-like leading edge of the jet continued rising without 
    rotation, leaving a dimming region behind before falling back to the solar surface. 
    The appearance/disappearance of dimming corresponded to the longitudinal 
    ascending/descending motions of jet. Cospatial H$\alpha$ surge and
    EUV dimming imply that the dimming resulted from the absorption of hot
    EUV emission by cool surge. The flare/jet event was caused by continuous
    magnetic cancellation before the start of flare. The jet was associated with
    the open magnetic fields at the edge of AR 11314.}
   {}


   \maketitle

\section{Introduction} \label{s-intro}

There are various jet-like activities in the solar atmosphere, such as 
spicules (De Pontieu et al. \cite{dep04}), chromospheric jets 
(Shibata et al. \cite{shi07}; Nishizuka et al. \cite{nis08}; 
Liu et al. \cite{liu09,liu11b}; Singh et al. \cite{sin12}), 
H$\alpha$ surges (Schmieder et al. \cite{sch95}; Zhang et al. \cite{zhang00};
Liu \& Kurokawa \cite{liu04}; Jiang et al. \cite{jiang07}), extreme-ultraviolet 
(EUV) jets (Chae et al. \cite{chae99}; Nistic{\`o} et al. \cite{nis09}; 
Moschou et al. \cite{mos12}), and soft X-ray (SXR) jets (Shibata et al. \cite{shi92}; 
Savcheva et al. \cite{sav07}). Most of the coronal jets 
seen in EUV and SXR are associated with flares, microflares, or bright points 
at their footpoints (Ji et al. \cite{ji08}; Zhang et al. \cite{zqm12}; 
Zhang \& Ji \cite{zqm13}). The typical height of jets is 10$-$400 Mm, 
the width is 5$-$100 Mm, the longitudinal 
velocity is 10$-$1000 km s$^{-1}$ with an average value of $\sim$200 km s$^{-1}$,
and the kinetic energy is in the order of 10$^{25}$--10$^{28}$ ergs 
(Shimojo et al. \cite{sho96}). Coronal jets are formed in coronal holes 
(Wang et al. \cite{wang98}; 
Cirtain et al. \cite{cir07}; Culhane \cite{cul07}; Patsourakos et al. \cite{pat08}) 
or at the edge of active regions (Kim et al. \cite{kim07}; Guo et al. \cite{guo13}) 
in the presence of open magnetic fields. Particles are 
accelerated and ejected into the interplanetary space during the 
reconnection, generating Type III radio bursts (Krucker et al. \cite{kru11}; 
Glesener et al. \cite{gle12}). Coronal jets
are believed to be heated by magnetic reconnection between emerging 
flux and the pre-existing magnetic fields (Yokoyama \& Shibata \cite{yoko96};
Moreno-Insertis et al. \cite{mor08}; T{\"o}r{\"o}k et al. \cite{tor09}; Jiang et al.
\cite{jiang12}; Moreno-Insertis \& Galsgaard \cite{mor13}; Pontin et al. \cite{pont13}). 
Moore et al. (\cite{moo10}) 
classified jets into two types: standard jets and blowout jets, the later of which 
features blowout eruption of the base arch\rq{}s core field.

Apart from the ordinary collimated motions, jets occasionally exhibit helical 
structure and untwisting motions (Liu et al. \cite{liu11a}; Shen et al. \cite{shen11}; 
Chen et al. \cite{chen12}; Hong et al. \cite{hong13}; Schmieder et al. \cite{sch13}). 
The amplitude of transverse
rotations ranges from 2 to 10 Mm. The periods are in the order of 4$-$9 min.
The transverse velocities (10$-$150 km s$^{-1}$) are slightly lower than their 
longitudinal velocities along the jet axis. The untwisting motions were previously
interpreted as the releasing of magnetic helicity during the reconnection 
between a twisted bipole and open fields (Shibata \& Uchida \cite{shi86}; 
Canfield et al. \cite{can96}; Jibben \& Canfield \cite{jib04}).
Pariat et al. (\cite{pari09,pari10}) numerically simulated the formation of 
untwisting jets as a result of continuous pumping of magnetic free 
energy and helicity into the corona from the photosphere, which is interpreted as 
upward propagation of torsional Alfv\'{e}n waves at a speed of hundreds of km s$^{-1}$.
Numerous flare-related jets and rotating jets have been observed. 
In this paper, we report the helical structure and swirling motion of a flare-related EUV jet
observed by the Atmospheric Imaging Assembly (AIA; Lemen et al. \cite{lem12}) aboard 
the Solar Dynamics Observatory (SDO) spacecraft. In \S\ref{s-data}, 
we describe the multiwavelength data analysis. The results are shown in \S\ref{s-result}. 
Discussion and summary are presented in \S\ref{s-disc}.

\section{Data analysis} \label{s-data}

SDO/AIA has unprecedentedly high cadence and resolution in the seven EUV wavelengths 
(94, 131, 171, 193, 211, 304, and 335 {\AA}). A C1.6 flare and the accompanying jet at the 
edge of NOAA active region 11314 was observed by AIA during 11:00$-$14:00 UT on 
2011 October 15. The full-disk level\_1 fits data were calibrated using the standard 
program {\it aia\_prep.pro} in the {\it Solar Software}. Image coalignments between the 
EUV passbands were performed using the cross-correlation method after selecting a 
bright feature (e.g., active region). The accuracy of coalignments is 1$\farcs$2. 
In order to investigate the three-dimensional (3D) structure of the jet, we checked 
data from the Extreme-Ultraviolet Imager (EUVI) in the Sun Earth Connection Coronal 
and Heliospheric Investigation (SECCHI; Howard et al. \cite{how08}) package aboard the 
Solar TErrestrial RElations Observatory (STEREO; Kaiser \cite{kai05}). The orbits of the 
twin satellites (A and B) drifted as time goes by, so that they had 
separation angles of $\sim$105$\degr$ and $\sim$99$\degr$ with the Sun-Earth 
connection on that day. Combined with SDO, the perspectives from STEREO  
provide excellent opportunity to get a full-view of solar eruptions. The
195 {\AA} raw data were calibrated by using the program {\it secchi\_prep.pro}. Deviation 
of the STEREO N-S direction from the solar rotation axis was corrected. After checking the
EUV observation during that time, we found that the jet was visible
only in the field-of-view of STEREO-B, but still partly blocked by the solar western limb. 
To explore the relationship between the hot
jet and cool surge, we examined the gound-based H$\alpha$ observation from the 
El Teide as a member of the Global Oscillation Network Group (GONG). The H$\alpha$
images were coaligned with the AIA 304 {\AA} images due to their lower formation heights.
Unfortunately, there was a data gap during 11:57$-$13:12 UT when the flare occurred.
The line-of-sight (LOS) magnetograms from the Helioseismic and Magnetic Imager 
(HMI; Scherrer et al. \cite{sch12}) aboard SDO were used to study the temporal
evolution of magnetic fields in the photosphere. The magnetograms were coaligned 
with the H$\alpha$ images according to the position of sunspot. Moreover, we carried out 
potential-field extrapolation to figure out the 3D magnetic configuration of the jet. The E-W 
and N-S scopes of the magnetogram for extrapolation were 886$\farcs$8 and 
706$\farcs$2 so that flux balance was guaranteed. SXR light curves from the GOES 
spacecraft were utilised to investigate the evolution of C1.6 flare. The observing
parameters are summarised in Table~\ref{table:1}, including the instruments, wavelengths
($\lambda$), observing times, cadences, and pixel sizes.

\begin{table}
\caption{Description of the Observational Parameters} 
\label{table:1}
\centering
\begin{tabular}{l c c c c}
\hline\hline
Instrument & $\lambda$ & Time & Cadence & Pixel Size \\ 
           &     ({\AA})   & (UT)  &    (sec)    &  (arcsec)  \\
\hline
  AIA & 94$-$335 & 11:00$-$14:00 & 12 & 0.6 \\
  HMI &  $-$         & 11:00$-$14:00 & 45 & 0.6 \\
  EUVI & 195       & 11:00$-$14:00 & 300 & 1.6 \\
  EI Teide & 6563 & 11:10$-$11:57 & 60  & 1.0 \\
  EI Teide & 6563 & 13:12$-$14:00 & 60  & 1.0 \\
  GOES   & 0.5$-$4.0  & 11:00$-$14:00 & 3  & $-$ \\ 
  GOES   & 1.0$-$8.0  &  11:00$-$14:00 & 3  & $-$  \\
\hline
\end{tabular}
\end{table}

\begin{figure}
\centering
\includegraphics[width=12cm,clip=]{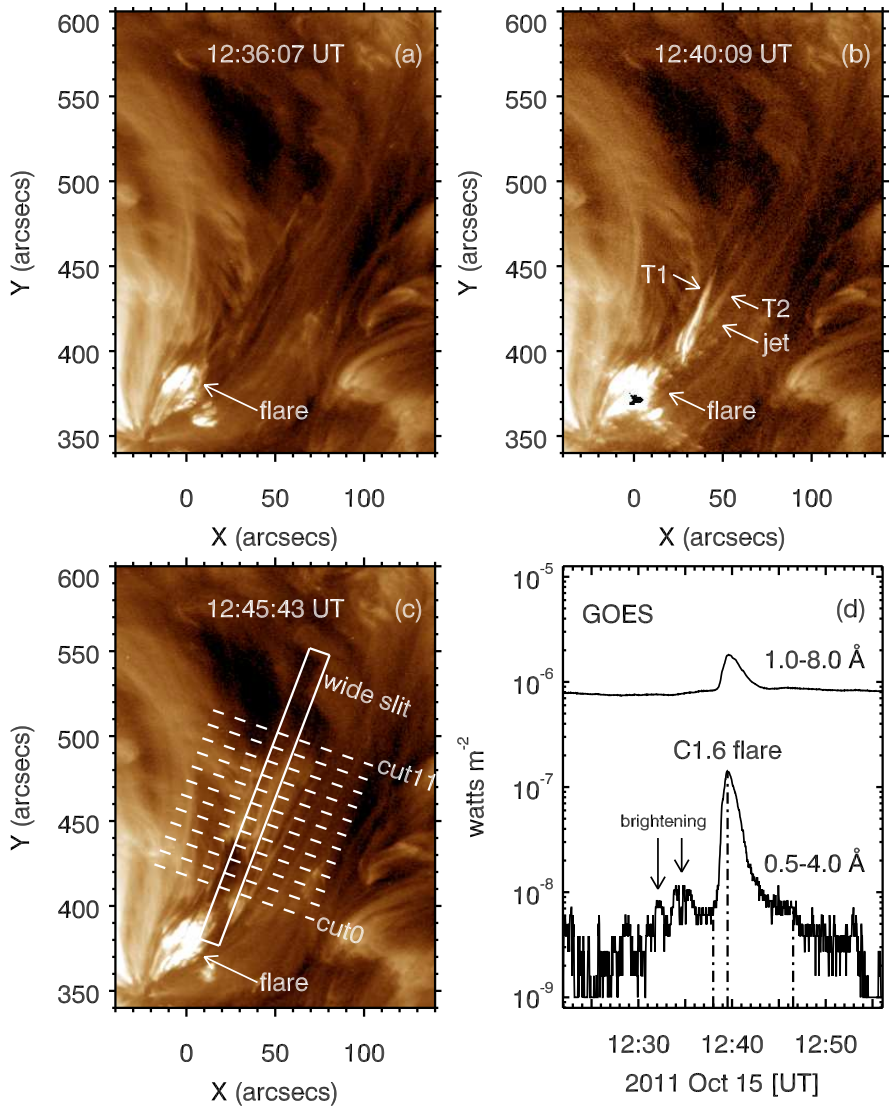}
\caption{{\bf (a)}$-${\bf (c)} Three snapshots of the 193 {\AA} images. The white arrows
point to the flare and jet. In panel {\bf (b)}, the two thick threads of jet are labeled with
``T1'' and ``T2''. In panel {\bf (c)}, the solid box and dashed lines represent the
wide slit and 12 narrow slits (cut0$-$cut11), respectively. {\bf (d)} GOES light curves in 
1$-$8 {\AA} (upper) and 0.5$-$4.0 {\AA} (lower), featuring the C1.6 flare and weak 
precursor brightenings. The dash-dotted lines denote the start (12:38:00 UT) and 
peak (12:39:30 UT) of the flare impulsive phase and end (12:46:30 UT) of the main phase.
The temporal evolution is shown in a movie with a larger field-of-view available in 
the online edition.}
\label{fig1}
\end{figure}

\section{Results} \label{s-result}

In Fig.~\ref{fig1}, the three snapshots of AIA 193 {\AA} images represent the
three phases of flare: pre-flare phase (a), impulsive phase (b), and main phase (c), 
respectively. In the pre-flare phase, there was very weak brightening at the flare site 
in the absence of jet. In the impulsive phase, the area and total emission of flare quickly
increased and reached maximum. The helical structure of jet was composed of two 
interwinding threads (``T1'' and ``T2'') that started counter-clockwise rotation 
perpendicular to its axis. 
In the main phase, the total emission of flare decreased significantly. The two compact 
threads untwisted into many thin threads, implying the multistrand nature of jet. 
Afterwards, the jet experienced curtain-like eruption along its axis. To investigate the 
longitudinal and transverse motions, we extracted a wide slit along the axis and 12 narrow 
slits (from cut0 to cut11) perpendicular to the axis in Fig.~\ref{fig1}c. The 
wide slit (solid box) is 12$\arcsec$ in width and 182$\arcsec$ in length. The narrow 
slits (dashed lines) are 96$\arcsec$ in length. In Fig.~\ref{fig1}d, we plot the light 
curves in 1.0$-$8.0 {\AA} (upper) and 0.5$-$4.0 {\AA} (lower). The impulsive phase 
of flare started at 12:38:00 UT and peaked at 12:39:30 UT, and the main phase ended 
at 12:46:30 UT, which is marked by the dash-dotted lines.

\begin{figure}
\centering
\includegraphics[width=12cm,clip=]{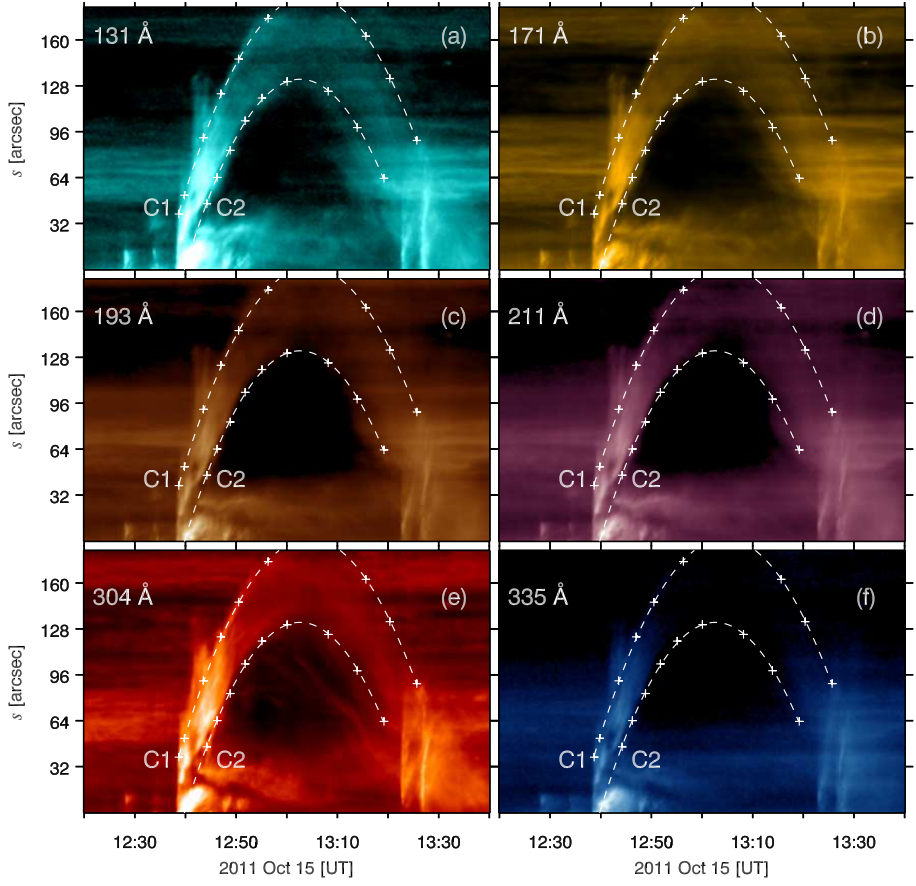}
\caption{{\bf (a)}$-${\bf (f)} 
Time-slice diagrams of the wide slit in six of the EUV passbands. $s=0\arcsec$ 
and 182$\arcsec$ in $y$-axis represent the bottom-left and top-right endpoints 
of the slit. The two dashed lines stand for the outer (C1) and inner (C2) 
boundaries of the parabolic trajectory.}
\label{fig2}
\end{figure}

In Fig.~\ref{fig2}, the time-slice diagrams of the wide slit in six of the EUV 
passbands (131, 171, 193, 211, 304, and 335 {\AA}) illustrate the longitudinal motion  
of jet along its axis. The jet underwent a parabolic trajectory during 
12:38$-$13:25 UT. We outline the outer and inner boundaries of the 
trajectory with two curves (C1 and C2) that are fitted with a quadratic function:

\begin{equation}
s=s_{0}+v_{\parallel}(t-t_{0})+\frac{1}{2}g_{\parallel}(t-t_{0})^{2},
\end{equation}
where $t_0=12$:38 UT, the initial velocity $v_{\parallel}=264$ and 244 km s$^{-1}$, 
the acceleration $g_{\parallel}=-103$ and $-$92 m s$^{-2}$ for C1 and C2, respectively. 
Combining the two curves, we obtain $v_{\parallel}=254\pm10$ km s$^{-1}$ and 
$g_{\parallel}=-97\pm5$ m s$^{-2}$.

\begin{figure}
\centering
\includegraphics[width=10cm,clip=]{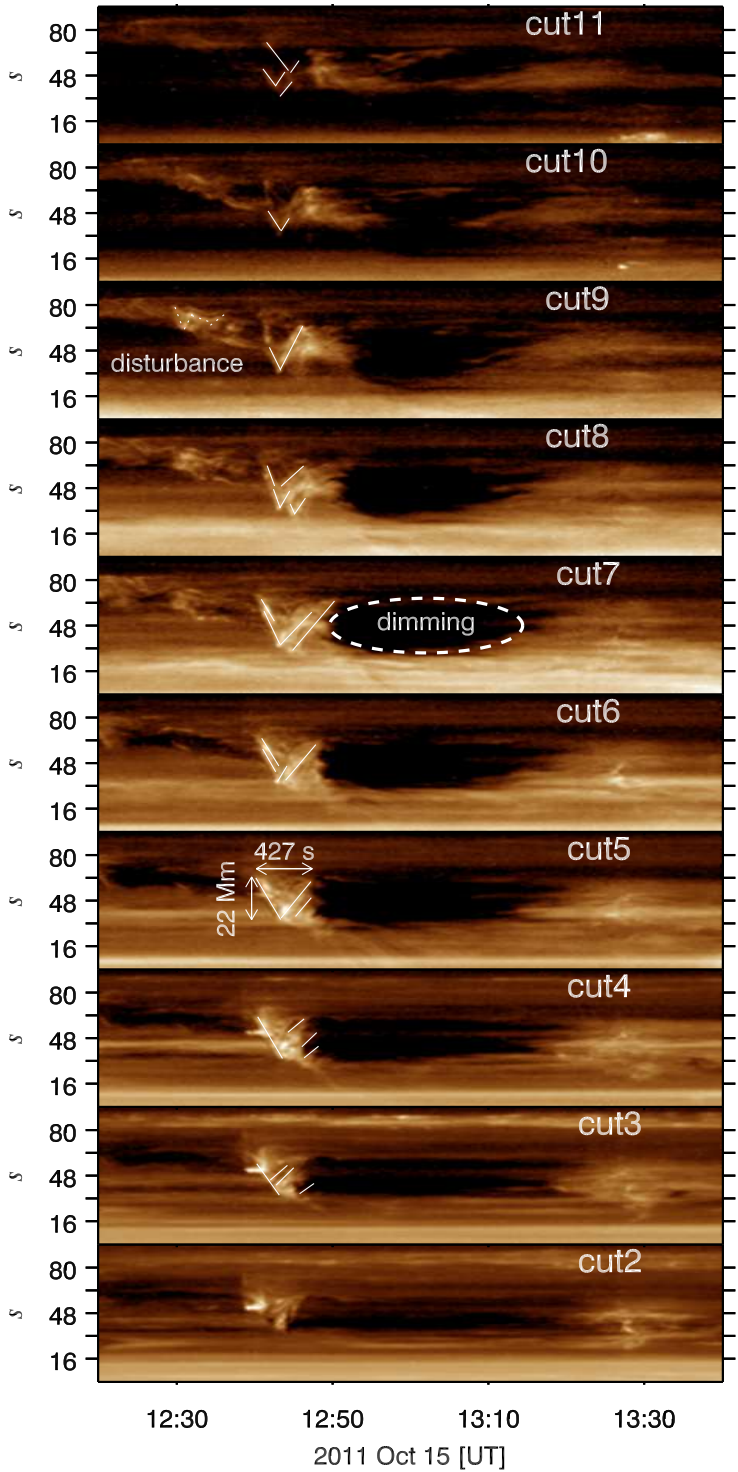}
\caption{Time-slice diagrams of the ten narrow slits (cut2$-$cut11) in 193 {\AA}.
$s=0\arcsec$ and 96$\arcsec$ in $y$-axis signify the top-left and bottom-right 
endpoints of slits. The short solid lines denote the 34 manually selected 
bright features signifying the leftwards and rightwards transverse rotation of
the jet. The short dotted lines in the third row denote the four manually selected 
bright features signifying the precursor disturbances adjacent to the jet axis.
The dashed ellipse in the fifth panel outlines the boundary of dimming. The 
horizontal and vertical double-headed arrows in the seventh panel measure 
the period (427 s) and double amplitude (22 Mm) of rotation.}
\label{fig3}
\end{figure}

\begin{figure}
\centering
\includegraphics[width=12cm,clip=]{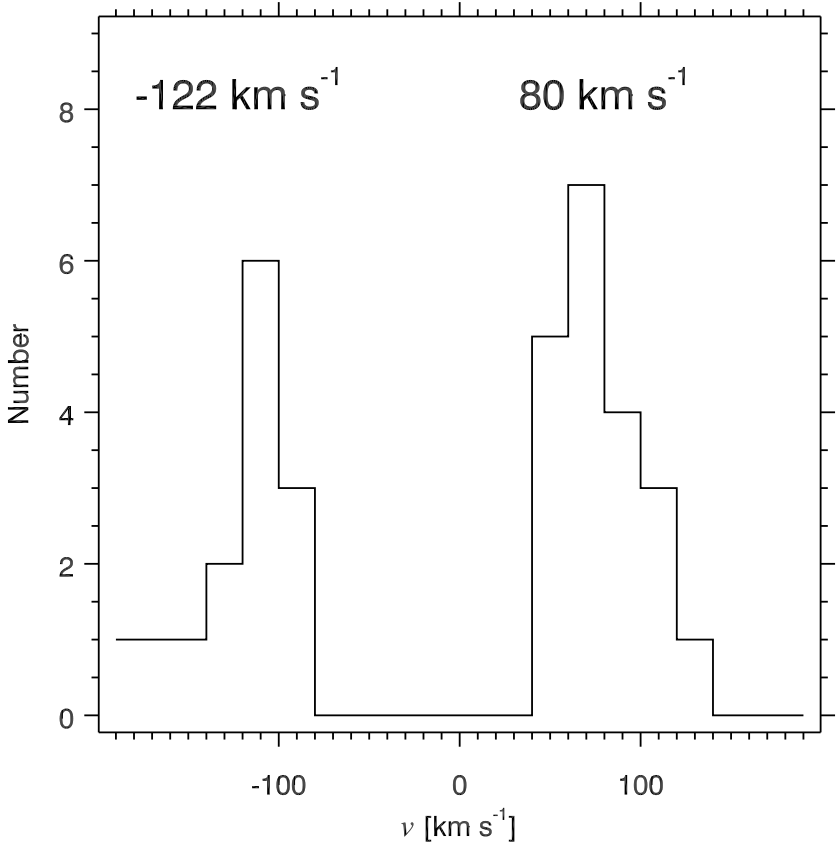}
\caption{Histogram of the transverse velocities of jet rotation. 
Negative/positive values suggest leftwards/rightwards motions.
The average leftward speed (-122 km s$^{-1}$) and average 
rightward speed (80 km s$^{-1}$) of the jet rotation are labeled.}
\label{fig4}
\end{figure}

Figure~\ref{fig3} displays the time-slice diagrams of 10 narrow slits (cut2$-$cut11) 
in 193 {\AA}. $s$ in $y$-axis of each panel signifies the distance from the 
top-left endpoints of slits. The middle of $y$-axis ($s=48\arcsec$) corresponds 
to the position of jet axis. It is clear that the jet rotated leftwards during 
$\sim$12:40$-$12:44 UT, which is illustrated by the solid oblique lines whose slopes
stand for the transverse velocities. The jet rotated rightwards during 
$\sim$12:44$-$12:48 UT, which is also illustrated by the oblique lines. We calculated
the jet velocities from the 34 labeled lines and show the histogram in Fig.~\ref{fig4}
where negative/positive values represent leftwards/rightwards rotation. It is revealed
that the rotation slowed down from an average leftward speed of 
$\sim$122 km s$^{-1}$ to an average rightward speed of $\sim$80 km s$^{-1}$.
The rotation lasted for only 1 cycle whose period (427 s) and twofold amplitude 
(22 Mm) are marked in the time-slice diagram of cut5, i.e., the seventh row of
Fig.~\ref{fig3}. The near-simultaneous
reversals of direction around 12:44 UT indicate that the whole jet rotated in phase.

After the transverse rotation, the helical jet became curtain-like and continued rising in the
longitudinal direction, leaving behind a dimming, which is manifested by the central dark regions
in all panels of Fig.~\ref{fig3}. We label the dimming region with a dashed ellipse in the fifth
row of Fig.~\ref{fig3}. The appearances of dimming occurred successively from the lower to 
upper parts of jet. The disappearances of dimming took place successively, however, from 
the upper to lower parts of jet. Combining with the longitudinal evolutions of the jet in Fig.~\ref{fig2}, 
we found that the appearance/disappearance of dimming coincided with the longitudinal 
ascending/descending motions.

\begin{figure}
\centering
\includegraphics[width=12cm,clip=]{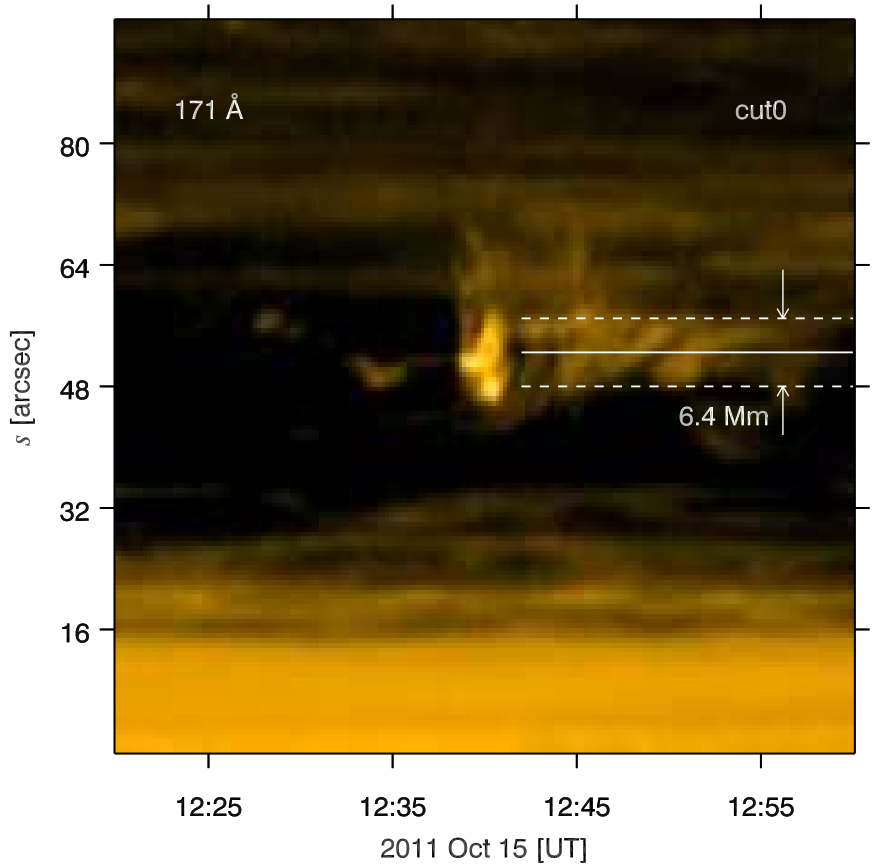}
\caption{Time-slice diagram of cut0 near the bottom of jet in 171 {\AA}. 
The solid and dashed lines mark the center and boundaries of the helical 
feature that indicates swirling motion of the jet. The distance between the
dashed lines (6.4 Mm) equals to double amplitude of jet rotation at
cut0.}
\label{fig5}
\end{figure}

Figure~\ref{fig5} shows the time-slice diagram of cut0 that is close to the base of 
jet. The helical feature in the diagram suggests transverse rotation during 
$\sim$12:42$-$12:55 UT. According to the handedness of feature, we conclude that
the jet rotated counter-clockwise with respect to its axis. We extracted the intensity 
profile along the center of the helical pattern, which is labeled with a solid line bounded 
by two dashed lines. The distance (6.4 Mm) between the dashed lines equals to 
double amplitude of rotation at cut0.

\begin{figure}
\centering
\includegraphics[width=12cm,clip=]{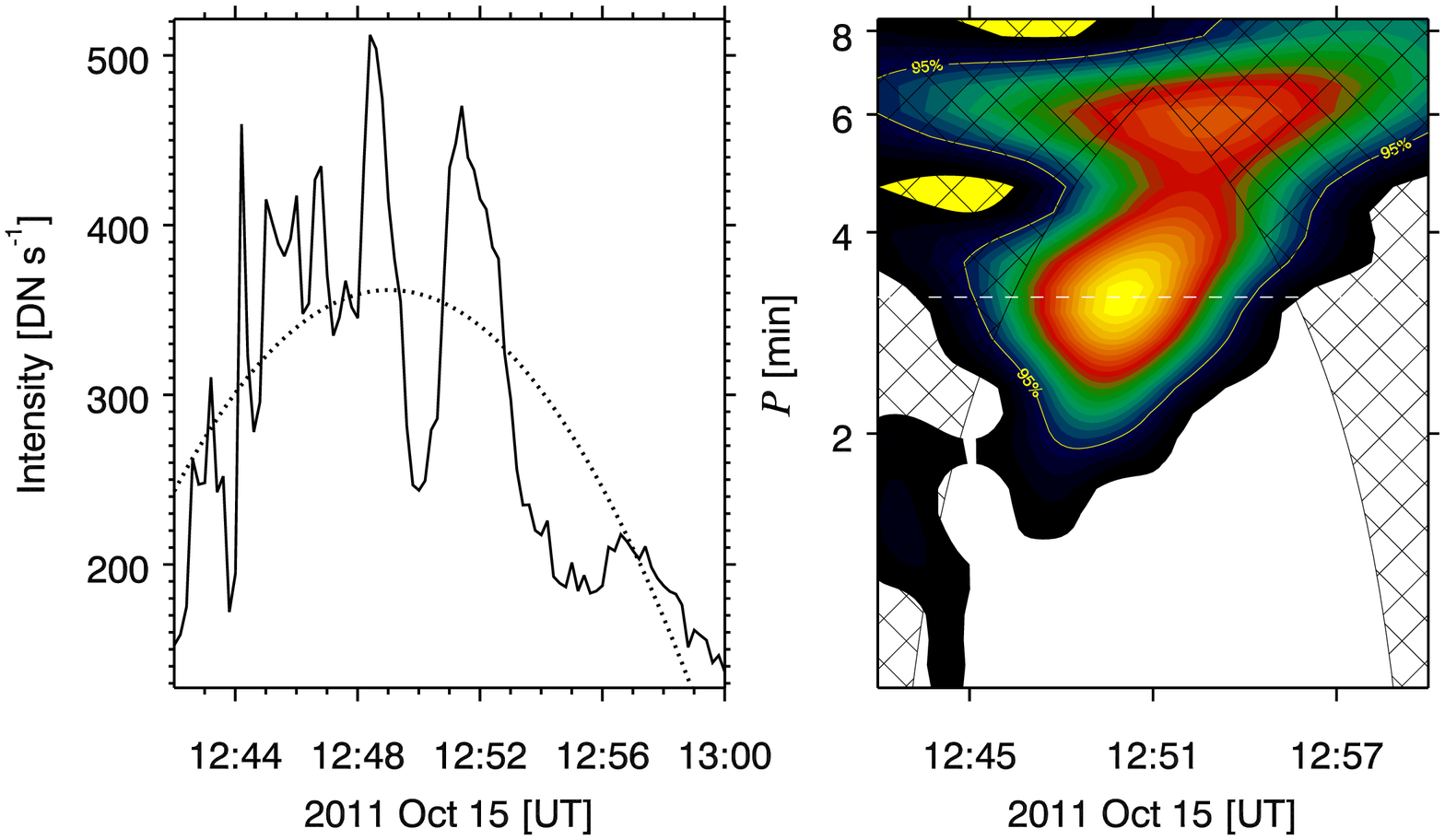}
\caption{{\it Left panel}: Intensity profile (solid) and its fitted trend (dotted) along the
solid line in Fig.~\ref{fig5}. 
{\it Right panel}: Wavelet transform of the trend-removed intensity profile. The white 
dashed line signifies the period ($\sim$192 s) with maximum power.}
\label{fig6}
\end{figure}

In the left panel of Fig.~\ref{fig6}, the intensity profile and its quadratic-fitted 
trend are plotted in solid and dotted lines. It is evident that the intensity 
profile presents quasi-periodic oscillation. The result of wavelet transform of the 
trend-removed profile is shown in the right panel where bright yellow region at the 
center suggests quasi-periodicity of the curve with period of $\sim$192 s. However, 
considering that the jet was composed of two compact interwinding threads at its 
bottom (``T1'' and ``T2'' in Fig.~\ref{fig1}b), the true period of rotation should be 
$\sim$384 s, which is slightly shorter than the measured value of 427 s in the 
seventh panel of Fig.~\ref{fig3}, implying that the jet rotates faster at the lower
part than the upper part. The amplitude of rotation at the base of jet ($\sim$3.2 Mm) 
accounts for 30\% of the value at the upper part.

\begin{figure}
\centering
\includegraphics[width=12cm,clip=]{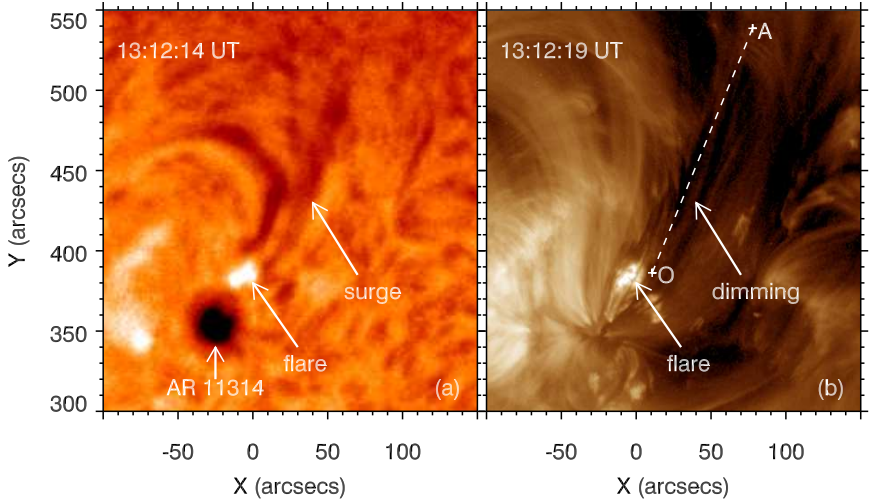}
\caption{{\it Left panel}: H$\alpha$ image at 13:12:14 UT featuring the AR 11314, 
flare, and cool surge as pointed by the white arrows. {\it Right panel}: AIA 193 {\AA} 
image at 13:12:19 UT featuring the flare and dimming following the leading
edge of jet as pointed by the white arrows. The dashed line (``OA'') is used for 
measuring the true height of jet.}
\label{fig7}
\end{figure}

Figure~\ref{fig7} shows the H$\alpha$ and AIA 193 {\AA} images at 
13:12 UT. It is clear that the cool surge seen in H$\alpha$ passband is cospatial 
with the dimming region following the jet leading edge. The surge was visible
until 13:17 UT. Unfortunately, there was a data gap before 13:12 UT. We are 
unable to investigate the temporal relation between the jet and surge during 
the flare.

\begin{figure}
\centering
\includegraphics[width=12cm,clip=]{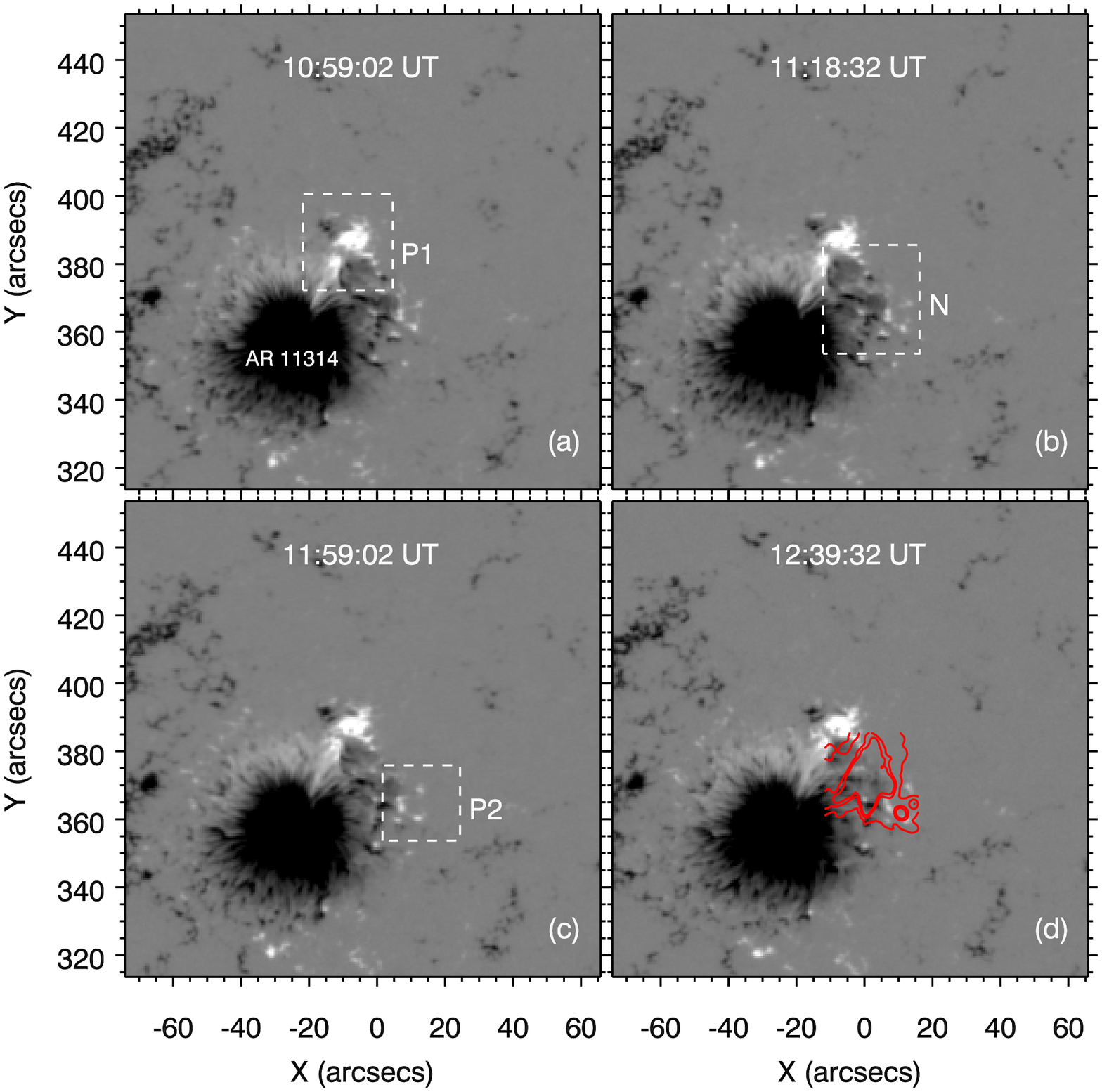}
\caption{HMI LOS magnetograms at 10:59:02 UT {\bf (a)}, 11:18:32 UT {\bf (b)}, 
11:59:02 UT {\bf (c)}, and 12:39:32 UT {\bf (d)}. ``P1'' and ``P2'' with positive 
polarity are included by the white dashed boxes in panels {\bf (a)} and {\bf (c)}.
``N'' with negative polarity is included by the white dashed box in panel
{\bf (b)}. The red lines in panel {\bf (d)} represent the intensity contour of flare
shown in the top-left panel of Fig.~\ref{fig10}.}
\label{fig8}
\end{figure}

\begin{figure}
\centering
\includegraphics[width=12cm,clip=]{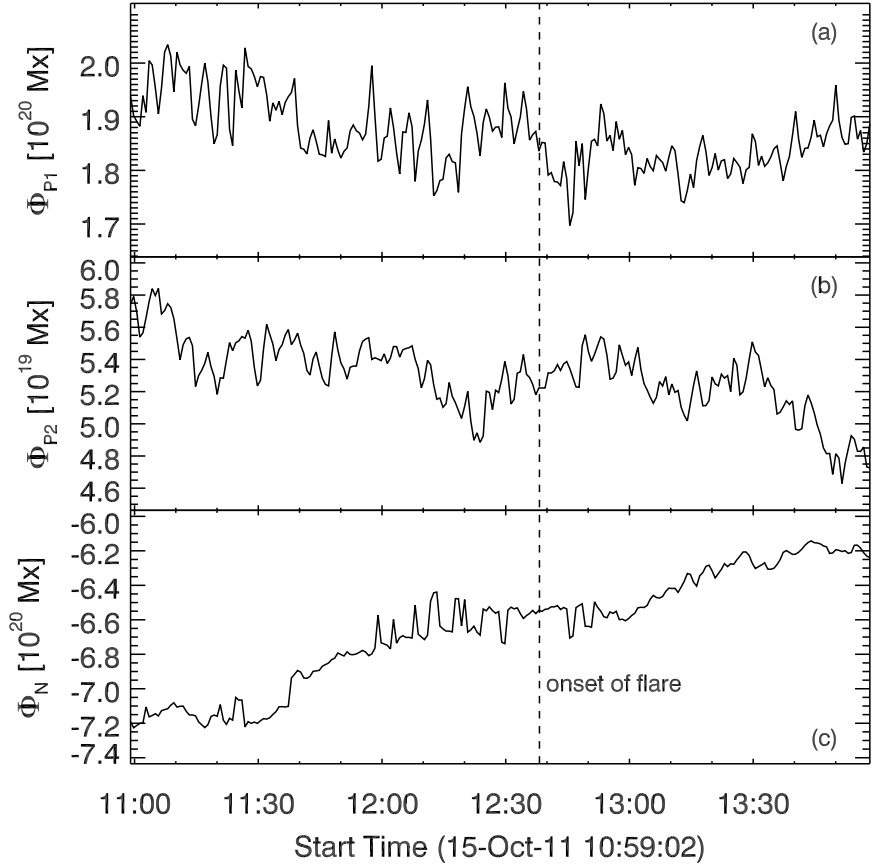}
\caption{Temporal evolutions of the magnetic fluxes of ``P1'' {\bf (a)}, ``P2'' {\bf (b)}, 
and ``N'' {\bf (c)}. The dashed line denotes the start of jet and flare impulsive phase.}
\label{fig9}
\end{figure}

The compact solar flare are usually associated with magnetic flux emergence or 
cancellation. To understand the cause of flare/jet event, we checked
the HMI LOS magnetograms. In Fig.~\ref{fig8}, we display four magnetograms around 
the sunspot of AR 11314. The flare took place near the north-west boundary of the strong 
sunspot with negative magnetic polarity (Fig.~\ref{fig8}d). We label the
negative-polarity region around the sunspot with ``N'' (Fig.~\ref{fig8}b) and the neighbouring
positive-polarity regions with ``P1'' (Fig.~\ref{fig8}a) and ``P2'' (Fig.~\ref{fig8}c). After viewing 
the movie of magnetograms, we discovered magnetic cancellation before the flare. The 
temporal evolutions of magnetic flux of the three regions are demonstrated in Fig.~\ref{fig9} 
where $\Phi_{\mathrm P1}$, $\Phi_{\mathrm P2}$, and $\Phi_{\mathrm N}$ represent the 
total positive magnetic flux of ``P1'', positive flux of ``P2'', and negative flux of ``N''. It is evident 
that all the three polarities underwent continuous cancellation with small-scale fluctuations 
before the onset of jet and flare impulsive phase at 12:38 UT. Using linear curve-fitting, we
derived the cancellation rate for the three polarities, being  
$d\Phi_{\mathrm P1}/dt=-1.1\times10^{17}$ Mx hr$^{-1}$, 
$d\Phi_{\mathrm P2}/dt=-4.6\times10^{16}$ Mx hr$^{-1}$, and
$d\Phi_{\mathrm N}/dt=-8.0\times10^{17}$ Mx hr$^{-1}$, respectively.
The magnetic fluxes of ``P2'' and ``N'' kept on cancelling after the flare, while the flux of ``P1'' 
fluctuated and kept near-constant. The observations of jets and surges as a result of flux
emergence and cancellation have been reported (Liu \& Kurokawa \cite{liu04}; 
Jiang et al. \cite{jiang07}; Chifor et al. \cite{chi08}).

\begin{figure}
 \centerline{\hspace*{0.005\textwidth}
             \includegraphics[width=0.45\textwidth,clip=]{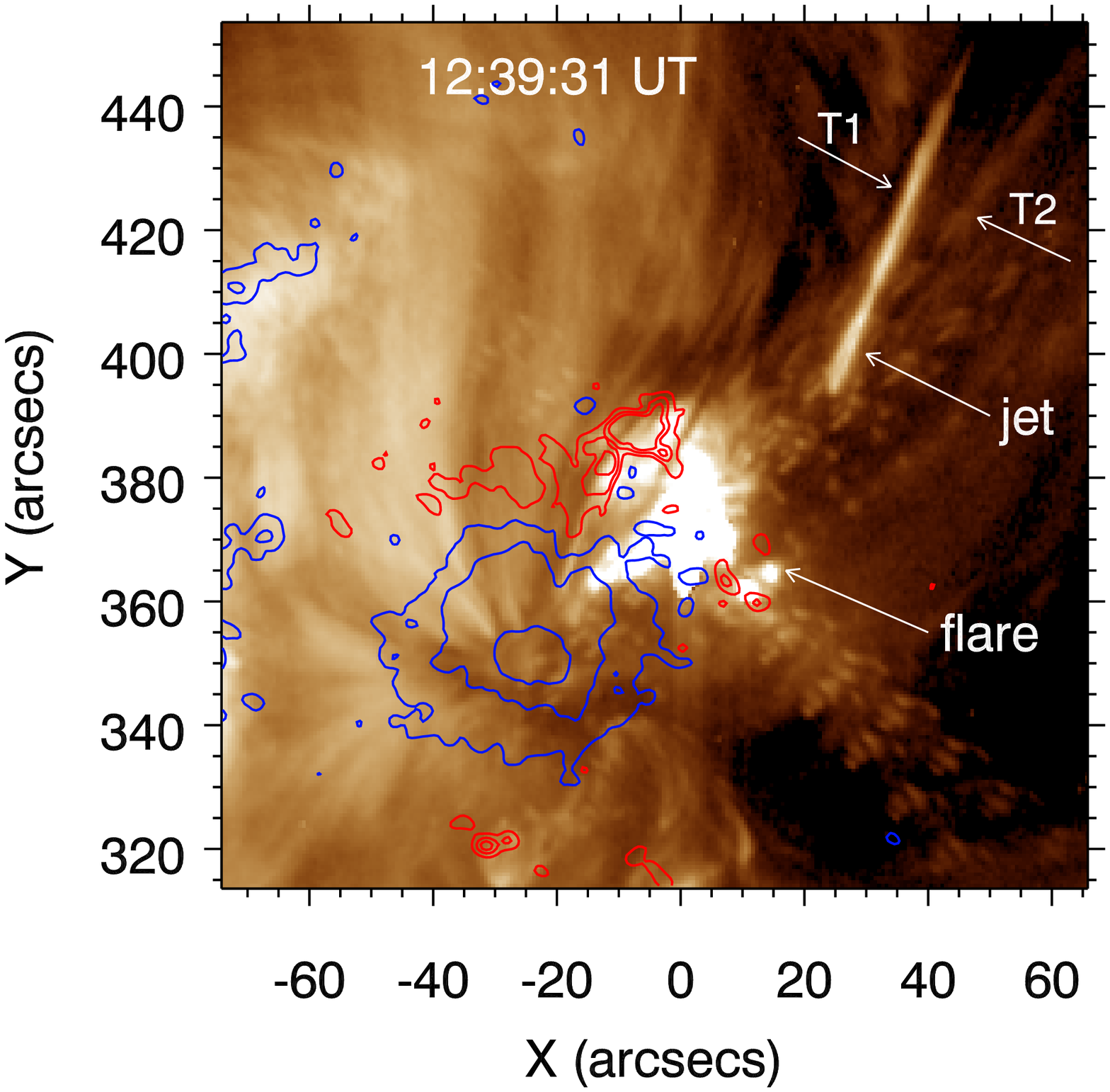}
             \hspace*{0.01\textwidth}
             \includegraphics[width=0.45\textwidth,clip=]{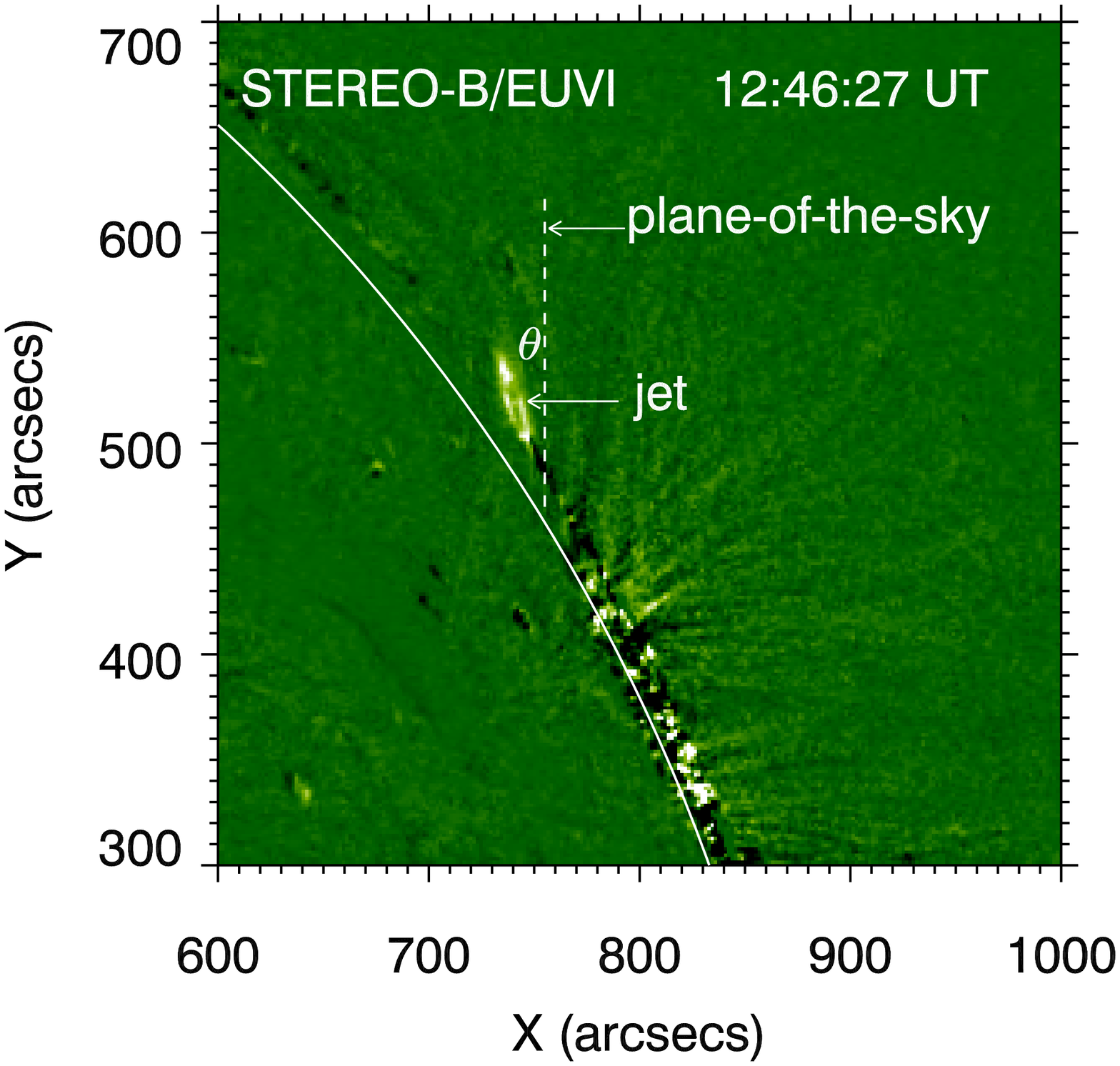}
            }

   \centerline{\hspace*{0.005\textwidth}
               \includegraphics[width=0.45\textwidth,clip=]{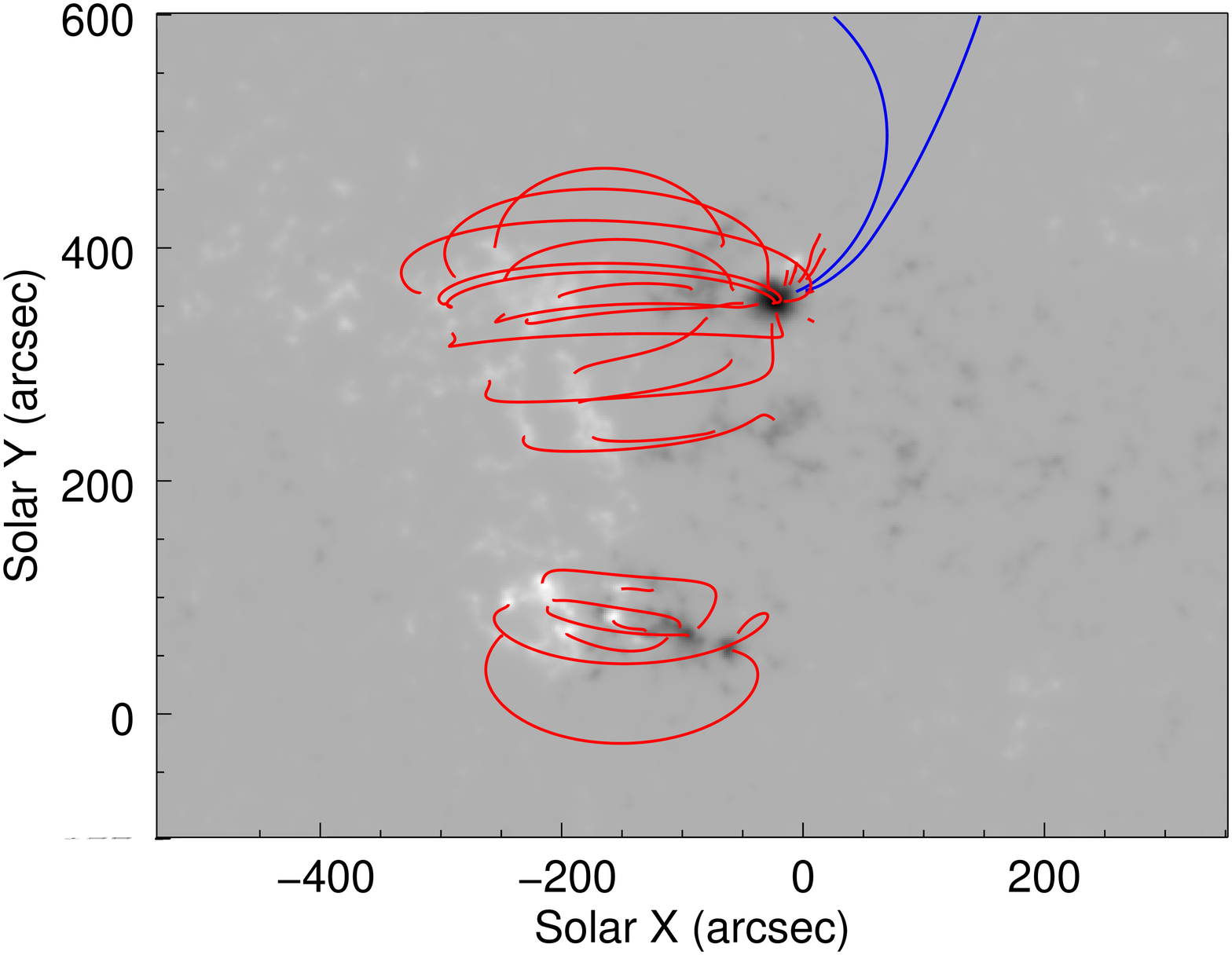}
               \hspace*{0.01\textwidth}
               \includegraphics[width=0.40\textwidth,clip=]{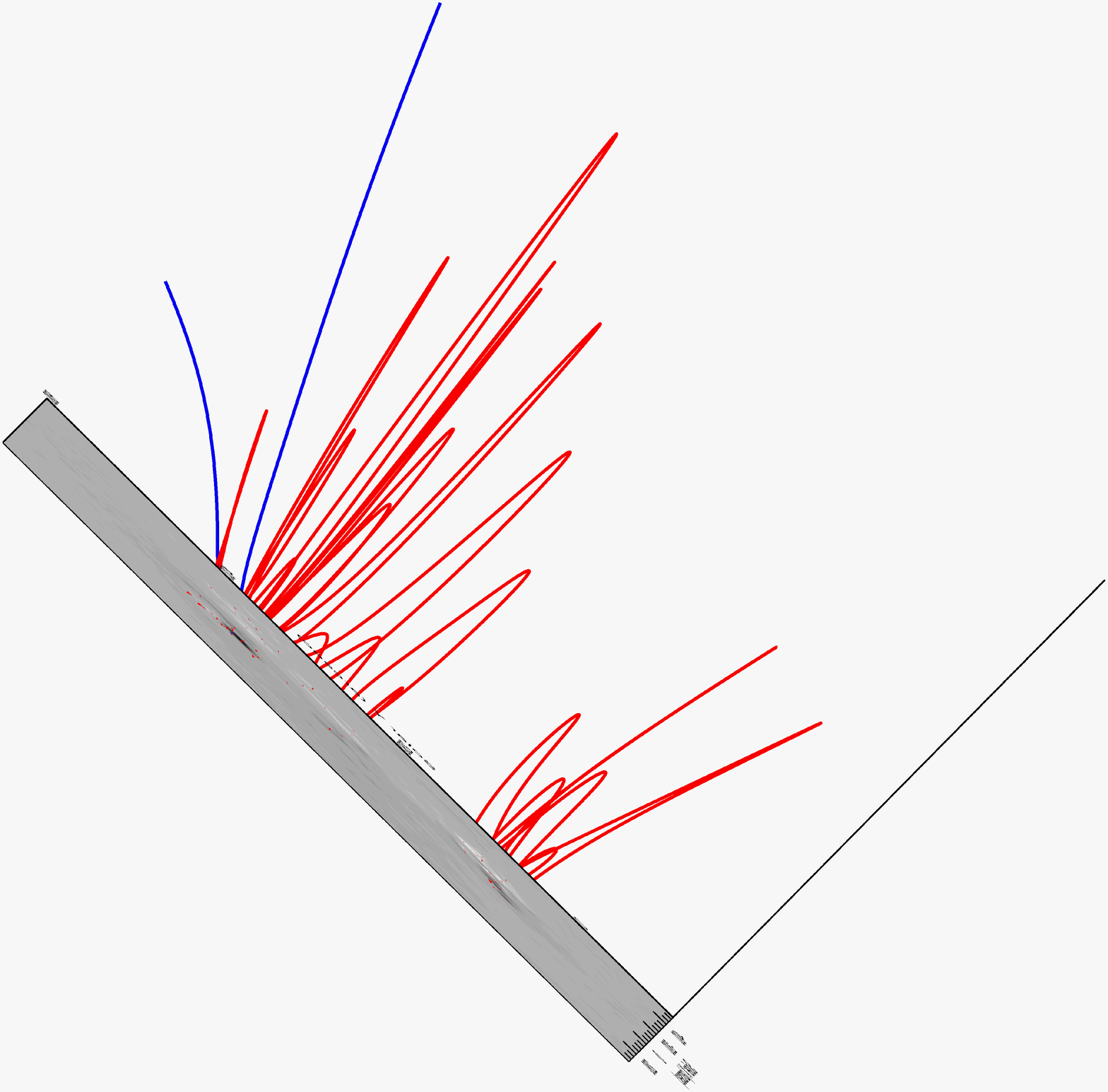}
              }
\caption{{\it Top left}: AIA 193 {\AA} image at 12:39:31 UT that features the jet threads 
(``T1'' and ``T2'') and flare as pointed by the white arrows. The red/blue lines stand for the 
contours of positive and negative polarities of the HMI LOS magnetogram at 12:39:32 UT. 
{\it Top right}: Running-difference image of STEREO-B/EUVI in 195 {\AA}
at 12:46:27 UT, showing the top segment of jet. The dashed line stands for the projection 
of plane-of-the-sky on the EUV image. The included angle between the jet and   
plane-of-the-sky is labeled with $\theta$.
{\it Bottom panels}: Top-view ({\it left}) and side-view ({\it right}) of the 3D magnetic field 
lines. The red/blue lines represent closed/open magnetic field lines.}
\label{fig10}
\end{figure}

To derive the 3D magnetic configuration of the flare-related jet, we performed 
potential-field extrapolation. In the top-left panel of Fig.~\ref{fig10}, we show the 
AIA 193 {\AA} image at 12:39:31 UT superposed by the HMI magnetogram 
contours at the same time where red/blue lines stand for positive/negative polarities. 
It features the bright compact flare and jet in the north-west direction. 
In the bottom-left panel of Fig.~\ref{fig10}, we display the top-view 
of magnetic configuration, where grayscale image represents the magnetogram and
the red/blue lines denote the closed/open magnetic field lines. It is obvious that the 
jet direction is well consistent with the orientation of projected open field lines.

In the top-right panel of Fig.~\ref{fig10}, we display the EUVI running-difference image 
at 12:46:27 UT in 195 {\AA}. Only the top segment of jet was observed while the lower 
segment of jet and flare were blocked. The included angle between the jet and the 
projection of plane-of-the-sky on the EUVI image (dashed line) is denoted with 
$\theta$ that measures 11.3$\degr$.
In the bottom-right panel of Fig.~\ref{fig10}, we demonstrate the side-view of the magnetic
field lines from STEREO-B viewpoint. It is obvious that the direction of jet is also 
consistent with the direction of open fields.

Since we have two perspectives to view the flare-related jet, one is from SDO/AIA,
the other is from STEREO-B/EUVI, we can take the projection effect into account to estimate
the true height and speed. In the right panel of Fig.~\ref{fig7}, we label the bottom and 
leading edge of jet with ``O'' and ``A'', whose distance equals to the apparent height 
of jet, i.e., 120.7 Mm. Considering the angular derivation ($\theta$) of jet from 
plane-of-the-sky, the true height and initial velocity are estimated to be 
122.7 Mm and 258$\pm10$ km s$^{-1}$.

\section{Discussion and summary} \label{s-disc}

The rising and subsequent falling of jets and surges along their axis have been extensively 
reported. Roy (\cite{roy73}) found that the acceleration of falling material of H$\alpha$ surges 
was less than the solar free fall acceleration ($g_{\odot}$). Liu et al. (\cite{liu09}) found that 
the effective gravitational acceleration of the \ion{Ca}{ii} H jet had a mean value of 141 
m s$^{-2}$. Shen et al. (\cite{shen11}) discovered an even smaller value (26 m s$^{-2}$)  
that was interpreted by the projection effect or damping effect of the background magnetic
fields. For our case, the parallel gravitational acceleration ($g_{\parallel}$) accounts for 
$\frac{1}{3}$$g_{\odot}-\frac{1}{2}$$g_{\odot}$ in the solar atmosphere. We propose 
that the Lorentz force of twisted jet threads (``T1'' and ``T2'' in Fig.~\ref{fig1}b) serves as an 
upward force against gravity.

Coronal mass ejections (CMEs; Chen \cite{chen11}; Cheng et al. \cite{cheng11}) are often 
accompanied by EUV waves and the following dimming that was interpreted by density 
depletion (Chen et al. \cite{chen02,chen10}). Such dimmings accompanying EUV jets have 
rarely been reported. Shen et al. (\cite{shen11}) observed a cavity obvious in 
304 {\AA} but undetectable in 193 {\AA}, suggesting the cool temperature of
cavity. Lee et al. (\cite{lee13}) discovered a fast jet-associated EUV dimming that is 
explained by Alfv\'{e}nic waves initiated by reconnection in the upper chromosphere.
In our study, the dimming following the leading edge of jet was detectable in all the EUV 
passbands of AIA. The appearance/disappearance of dimming coincided with the 
longitudinal rising/falling motions of jet. It is revealed in Fig.~\ref{fig7} that the dimming is 
cospatial with the H$\alpha$ surge, indicating that the dimming resulted from the 
absorption of hot EUV emission by the cool surge. The disappearance of surge at
$\sim$13:18 UT coincided with the disappearance of dimming around 13:20 UT.
Yokoyama \& Shibata (\cite{yoko96}) performed MHD numerical simulations to explain 
the generation of hot jet and cool surge during the magnetic reconnection between the
emerging magnetic flux and the pre-existing fields of opposite polarity. The jet and surge
are adjacent in the 2D simulation. However, if we observe the eruption from an 
appropriate viewpoint, the jet would be blocked by the foreground surge, leading to the
absorption of EUV emission by the H$\alpha$ surge at a lower speed.

The helical structure and untwisting motions of EUV jets have been extensively investigated 
since the launch of high-resolution space telescopes (e.g., SDO). Such motions
were often explained by the releasing of accumulated magnetic helicity into the upper 
solar atmosphere. The jet we observed underwent counter-clockwise rotation at the 
beginning of its eruption. The transverse velocities range from 40 to 200 km s$^{-1}$,
very close to the values reported by Shen et al. (\cite{shen11}).
The rotation slowed down from $\sim$122 km s$^{-1}$ in the initial phase to $\sim$80
km s$^{-1}$ in the later phase, which is in agreement with the cases of Liu et al. (\cite{liu09})
and Schmieder et al. (\cite{sch13}).
The rotation, however, lasted for only 1 cycle with period of $\sim$7 min, which is in the 
same order of magnitude as the previously reported values (Liu et al. \cite{liu09}; Chen et al. 
\cite{chen12}).

\begin{figure}
\centering
\includegraphics[width=12cm,clip=]{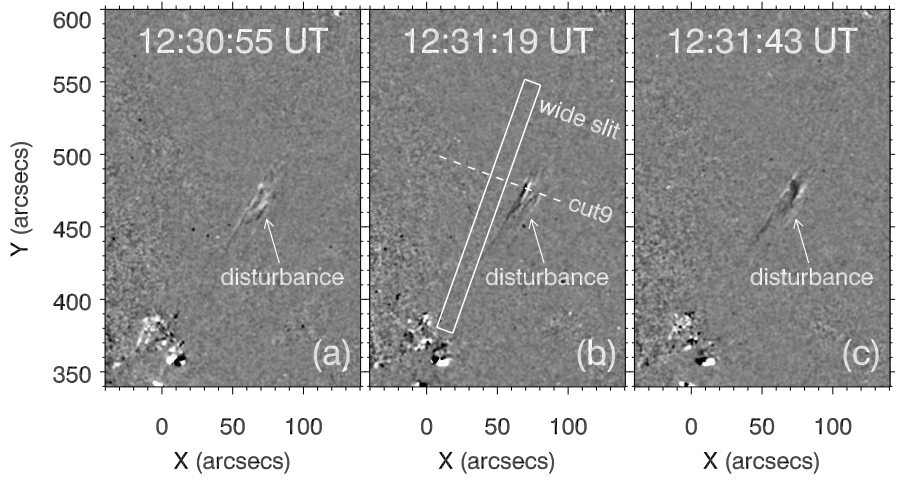}
\caption{Running-difference images in 193 {\AA} during 12:30$-$12:32 UT. White/black
color denotes intensity increase/decrease. The arrows point to the transverse disturbances 
adjacent to the jet axis prior to the flare.}
\label{fig11}
\end{figure}

Precursor disturbances before the transverse jet rotation were observed. Figure~\ref{fig11}
shows the running-difference images during 12:30$-$12:32 UT in 193 {\AA}, where 
white/black color represents intensity increase/decrease. Transverse disturbances 
adjacent to the jet axis are distinguishable in the images. In the time-slice diagram of cut9 
(third row of Fig.~\ref{fig3}), we label the disturbances with four dotted lines, the slope of 
which represent the velocities, being $-$175, 126, $-$24, 51 km s$^{-1}$, respectively. As 
to the cause of the disturbances, we noticed weak brightenings during 12:30$-$12:36 UT 
before the impulsive phase of flare in the 0.4$-$5.0 {\AA} light curve (Fig.~\ref{fig1}d).
Precursor brightenings, implying small-scale and weak release of magnetic  
energy prior to the impulsive phase of flares, have been observed 
(Harrison et al. \cite{hari85}; Chifor et al. \cite{chi07}), and are consistent with the 
mechanism of blowout jets (Moore et al. \cite{moo10}). 

In this paper, we report a flare-related jet observed by SDO/AIA on 2011 October 15.
The jet underwent longitudinal rising/falling along its axis with an initial velocity of 254$\pm10$ 
km s$^{-1}$. The effective gravitational acceleration was $-97\pm5$ m s$^{-2}$, well
below the free fall acceleration. The onset of jet eruption coincided with the beginning of 
impulsive phase of the adjacent C1.6 flare. At the beginning of its longitudinal eruption, the jet 
presented helical structure and counter-clockwise swirling motion that slowed down from 
an average of $\sim$122 km s$^{-1}$ in the initial stage to $\sim$80 km s$^{-1}$ in the 
later stage. The thick interwinding threads of jet untwisted into multiple thin threads 
during the rotation that lasted for only 1 cycle with period of $\sim$7 min and amplitude
that increases from $\sim$3.2 Mm at the lower part to $\sim$11 Mm at the upper part.
Afterwards, a dimming region appeared following the curtain-like leading edge of jet. 
The appearance/disappearance of dimming coincided with the ascending/descending 
motions. The cospatial EUV dimming and H$\alpha$ surge indicate that the dimming
resulted from the absorption of hot EUV emission by the cool surge. LOS mangetograms
from SDO/HMI show that the flare/jet event was caused by continuous magnetic
cancellation before the eruption. Potential-field extrapolation based on the magnetograms
reveals that the jet was associated with the open magnetic fields at the boundary of 
AR 11314. The true height and velocity of jet were estimated after considering the projection
effect from the two perspectives of SDO and STEREO-B.

\begin{acknowledgements}
The authors are grateful to the referee for enlightening comments. Q.M.Z 
acknowledges J. Q. Sun, P. F. Chen, Y. Guo, R. Erd\'{e}lyi, G. Verth, Y. N. Su, 
and the solar 
physics group in Purple Mountain Observatory for valuable discussions and 
suggestions. Q.M.Z is also thankful for the Department of Applied Mathematics 
in the University of Sheffield for their hospitality during his visit. SDO is a mission 
of NASA\rq{}s Living With a Star Program. AIA and HMI data are courtesy of the 
NASA/SDO science teams. STEREO/SECCHI data are provided by a 
consortium of US, UK, Germany, Belgium, and France. The Global Oscillation
Network Group (GONG) Program is managed by the National Solar Observatory. 
This work is supported by 973 program under grant 2011CB811402 and
by NSFC 11303101, 11333009, 11173062 and 11221063.
\end{acknowledgements}


\begin{thebibliography}{} 
\bibitem[1996]{can96} Canfield, R.~C., 
Reardon, K.~P., Leka, K.~D., et al.\ 1996, \apj, 464, 1016
\bibitem[1999]{chae99} Chae, J., Qiu, J., Wang, 
H., \& Goode, P.~R.\ 1999, \apjl, 513, L75 
\bibitem[2010]{chen10} Chen, F., Ding, M.~D., 
\& Chen, P.~F.\ 2010, \apj, 720, 1254
\bibitem[2012]{chen12} Chen, H. D., Zhang, J., \& Ma, S. L.\ 2012, 
Research in Astronomy and Astrophysics, 12, 573
\bibitem[2011]{chen11} Chen, P.~F.\ 2011, Living Reviews 
in Solar Physics, 8, 1
\bibitem[2002]{chen02} Chen, P.~F., Wu, S.~T., 
Shibata, K., \& Fang, C.\ 2002, \apjl, 572, L99
\bibitem[2011]{cheng11} Cheng, X., Zhang, J., 
Liu, Y., \& Ding, M.~D.\ 2011, \apjl, 732, L25
\bibitem[2007]{chi07} Chifor, C., Tripathi, D., Mason, H.~E., 
\& Dennis, B.~R.\ 2007, \aap, 472, 967
\bibitem[2008]{chi08} Chifor, C., Isobe, H., 
Mason, H.~E., et al.\ 2008, \aap, 491, 279
\bibitem[2007]{cir07} Cirtain, J.~W., Golub, 
L., Lundquist, L., et al.\ 2007, Science, 318, 1580
\bibitem[2007]{cul07} Culhane, L., Harra, 
L.~K., Baker, D., et al.\ 2007, \pasj, 59, 751
\bibitem[2004]{dep04} De Pontieu, B., 
Erd{\'e}lyi, R., \& James, S.~P.\ 2004, \nat, 430, 536
\bibitem[2012]{gle12} Glesener, L., Krucker, 
S., \& Lin, R.~P.\ 2012, \apj, 754, 9
\bibitem[2013]{guo13} Guo, Y., D{\'e}moulin, P., 
Schmieder, B., et al.\ 2013, \aap, 555, A19
\bibitem[1985]{hari85} Harrison, R.~A., Waggett, P.~W., 
Bentley, R.~D., et al.\ 1985, \solphys, 97, 387
\bibitem[2013]{hong13} Hong, J.-C., Jiang, Y.-C., Yang, J.-Y., et al.\ 2013, 
Research in Astronomy and Astrophysics, 13, 253
\bibitem[2008]{how08} Howard, R.~A., Moses, 
J.~D., Vourlidas, A., et al.\ 2008, \ssr, 136, 67
\bibitem[2008]{ji08} Ji, H., Wang, H., Liu, C., 
\& Dennis, B.~R.\ 2008, \apj, 680, 734
\bibitem[2012]{jiang12} Jiang, R.-L., Fang, C., 
\& Chen, P.-F.\ 2012, \apj, 751, 152
\bibitem[2007]{jiang07} Jiang, Y.~C., Chen, H.~D., 
Li, K.~J., Shen, Y.~D., \& Yang, L.~H.\ 2007, \aap, 469, 331
\bibitem[2004]{jib04} Jibben, P., \& Canfield, R.~C.\ 2004, \apj, 610, 1129
\bibitem[2005]{kai05} Kaiser, M.~L.\ 2005, Advances 
in Space Research, 36, 1483
\bibitem[2007]{kim07} Kim, Y.-H., Moon, Y.-J., 
Park, Y.-D., et al.\ 2007, \pasj, 59, 763
\bibitem[2011]{kru11} Krucker, S., Kontar, E.~P., 
Christe, S., Glesener, L., \& Lin, R.~P.\ 2011, \apj, 742, 82
\bibitem[2013]{lee13} Lee, K.-S., Innes, D.~E., 
Moon, Y.-J., et al.\ 2013, \apj, 766, 1 
\bibitem[2012]{lem12} Lemen, J.~R., Title, 
A.~M., Akin, D.~J., et al.\ 2012, \solphys, 275, 17
\bibitem[2011a]{liu11a} Liu, C., Deng, N., Liu, R., 
et al.\ 2011a, \apjl, 735, L18
\bibitem[2009]{liu09} Liu, W., Berger, T.~E., 
Title, A.~M., \& Tarbell, T.~D.\ 2009, \apjl, 707, L37
\bibitem[2011b]{liu11b} Liu, W., Berger, T.~E., 
Title, A.~M., Tarbell, T.~D., \& Low, B.~C.\ 2011b, \apj, 728, 103
\bibitem[2004]{liu04} Liu, Y., 
\& Kurokawa, H.\ 2004, \apj, 610, 1136
\bibitem[2010]{moo10} Moore, R.~L., Cirtain, 
J.~W., Sterling, A.~C., \& Falconer, D.~A.\ 2010, \apj, 720, 757
\bibitem[2008]{mor08} Moreno-Insertis, F., 
Galsgaard, K., \& Ugarte-Urra, I.\ 2008, \apjl, 673, L211
\bibitem[2013]{mor13} Moreno-Insertis, F., \& Galsgaard, K.\ 2013, 
\apj, 771, 20
\bibitem[2012]{mos12} Moschou, S.~P., Tsinganos, 
K., Vourlidas, A., \& Archontis, V.\ 2012, \solphys, 310
\bibitem[2008]{nis08} Nishizuka, N., 
Shimizu, M., Nakamura, T., et al.\ 2008, \apjl, 683, L83
\bibitem[2009]{nis09} Nistic{\`o}, G., 
Bothmer, V., Patsourakos, S., \& Zimbardo, G.\ 2009, \solphys, 259, 87
\bibitem[2009]{pari09} Pariat, E., Antiochos, 
S.~K., \& DeVore, C.~R.\ 2009, \apj, 691, 61
\bibitem[2010]{pari10} Pariat, E., Antiochos, 
S.~K., \& DeVore, C.~R.\ 2010, \apj, 714, 1762
\bibitem[2008]{pat08} Patsourakos, S., 
Pariat, E., Vourlidas, A., Antiochos, S.~K., 
\& Wuelser, J.~P.\ 2008, \apjl, 680, L73
\bibitem[2013]{pont13} Pontin, D.~I., Priest, 
E.~R., \& Galsgaard, K.\ 2013, \apj, 774, 154
\bibitem[1973]{roy73} Roy, J.-R.\ 1973, \solphys, 32, 139
\bibitem[2007]{sav07} Savcheva, A., Cirtain, 
J., Deluca, E.~E., et al.\ 2007, \pasj, 59, 771
\bibitem[2012]{sch12} Scherrer, P.~H., 
Schou, J., Bush, R.~I., et al.\ 2012, \solphys, 275, 207
\bibitem[1995]{sch95} Schmieder, B., Shibata, K., 
van Driel-Gesztelyi, L., \& Freeland, S.\ 1995, \solphys, 156, 245
\bibitem[2013]{sch13} Schmieder, B., Guo, Y., 
Moreno-Insertis, F., et al.\ 2013, \aap, 559, A1
\bibitem[2011]{shen11} Shen, Y., Liu, Y., Su, J., 
\& Ibrahim, A.\ 2011, \apjl, 735, L43
\bibitem[1986]{shi86} Shibata, K., 
\& Uchida, Y.\ 1986, \solphys, 103, 299
\bibitem[1992]{shi92} Shibata, K., Ishido, 
Y., Acton, L.~W., et al.\ 1992, \pasj, 44, L173
\bibitem[2007]{shi07} Shibata, K., Nakamura, 
T., Matsumoto, T., et al.\ 2007, Science, 318, 1591
\bibitem[1996]{sho96} Shimojo, M., Hashimoto, 
S., Shibata, K., et al.\ 1996, \pasj, 48, 123
\bibitem[2012]{sin12} Singh, K.~A.~P., Isobe, H., 
Nishizuka, N., Nishida, K., \& Shibata, K.\ 2012, \apj, 759, 33
\bibitem[2009]{tor09} T{\"o}r{\"o}k, 
T., Aulanier, G., Schmieder, B., Reeves, K.~K., 
\& Golub, L.\ 2009, \apj, 704, 485
\bibitem[1998]{wang98} Wang, Y.-M., Sheeley, 
N.~R., Jr., Socker, D.~G., et al.\ 1998, \apj, 508, 899
\bibitem[1996]{yoko96} Yokoyama, T., 
\& Shibata, K.\ 1996, \pasj, 48, 353
\bibitem[2000]{zhang00} Zhang, J., Wang, J., \& Liu, Y.\ 2000, 
\aap, 361, 759
\bibitem[2012]{zqm12} Zhang, Q.~M., Chen, 
P.~F., Guo, Y., Fang, C., \& Ding, M.~D.\ 2012, \apj, 746, 19
\bibitem[2013]{zqm13} Zhang, Q.~M., \& Ji, H.~S.\ 2013, 
\aap, 557, L5
\end{thebibliography}
\end{document}